\documentclass[conference]{IEEEtran}
\usepackage[]{footmisc}
\usepackage{graphicx}
\usepackage{amsmath}
\usepackage{algorithm}
\usepackage[noend]{algpseudocode}
\usepackage{array}
\usepackage{color}
\usepackage{colortbl}
\usepackage{multirow}
\usepackage{microtype}
\usepackage{enumitem}
\usepackage{etoolbox}
\usepackage[]{footmisc}
\usepackage{amssymb}

\newcommand{\etal}{\textit{et al. }}

\title{DANTE: A Framework for Mining and Monitoring Darknet Traffic}

\begin{document}

\author{
\IEEEauthorblockN{Dvir Cohen, Yisroel Mirsky, Yuval Elovici, Rami Puzis, and Asaf Shabtai \\
Department of Software and Information Systems Engineering \\
Ben Gurion University of the Negev \\
\{dvircohe,yisroel\}@post.bgu.ac.il, \{elovici,puzis,shabtaia\}@bgu.ac.il}
\\

\IEEEauthorblockN{Manuel Kamp, Tobias Martin \\
Deutsche Telekom AG \\
manuel.kamp@t-systems.com, tobias.martin@telekom.de}

}

\maketitle

\begin{abstract}
Trillions of network packets are sent over the Internet to destinations which do not exist. 
This `darknet' traffic captures the activity of botnets and other malicious campaigns aiming to discover and compromise devices around the world. 
In order to mine threat intelligence from this data, one must be able to handle large streams of logs and represent the traffic patterns in a meaningful way. 
However, by observing how network ports (services) are used, it is possible to capture the intent of each transmission.
In this paper, we present DANTE: a framework and algorithm for mining darknet traffic. 
DANTE learns the meaning of targeted network ports by applying Word2Vec to observed port sequences. 
Then, when a host sends a new sequence, DANTE represents the transmission as the average embedding of the ports found that sequence. 
Finally, DANTE uses a novel and incremental time-series cluster tracking algorithm on observed sequences to detect recurring behaviors and new emerging threats.
To evaluate the system, we ran DANTE on a full year of darknet traffic (over three Tera-Bytes) collected by the largest telecommunications provider in Europe, Deutsche Telekom and analyzed the results. DANTE discovered 1,177 new emerging threats and was able to track malicious campaigns over time. We also compared DANTE to the current best approach and found DANTE to be more practical and effective at detecting darknet traffic patterns.

\end{abstract}


\section{\label{sec:introduction}Introduction}

One way for Internet service providers (ISP) to obtain meaningful and actionable insights on malicious Internet campaigns,  is to analyze traffic arriving at a subset of unassigned IP addresses.
These IP addresses are sometimes referred to as a darknet ~\cite{bailey2005internet, bailey2006practical}. In this paper, the term `darknet' should not be confused with anonymous communication networks such as Tor.

\begin{figure}[t]
\centering
\includegraphics[width=\columnwidth]{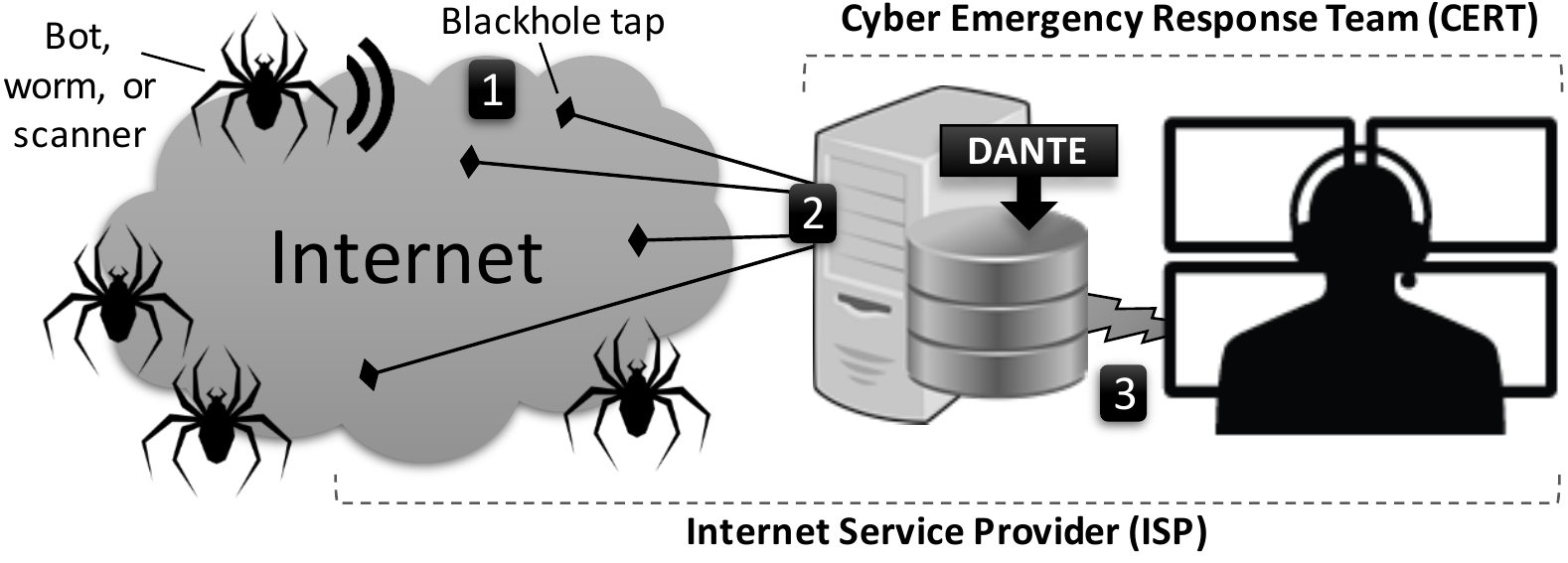}
\caption{Topological view of darknet analysis maintained by an ISP: (1) packets sent to non-existing IPs are captured at taps, then (2) logs are aggregated and sent to a cluster, and then (3) algorithms are used to generate emerging threat intelligence for the CERT.}
\label{fig:dante_figure}
\end{figure}

Darknet IP addresses are not associated with any registered host or services. Therefore, they are similar to honeypots~\cite{{mairh2011honeypot, bringer2012honeypot}} in that any incoming packets can be automatically considered unwanted and non-productive.
Previous studies have shown that packets sent to darknet IP addresses are usually the result of network probing/scanning, worm propagation, and a DDoS attacks ~\cite{Pang2016, Wustrow2010}. 
Therefore, darknet data can be used by an ISP's cyber emergency response team (CERT) to infer threat intelligence related to ongoing malicious activities or new emerging attacks~\cite{ban2012} (see Figure \ref{fig:dante_figure}).

A great advantage to this approach is that darknet taps are easily deployed, inexpensive to implement, and can collect vast amounts of useful data.
However, obtaining threat intelligence from darknet traffic is a challenging task for three reasons: 
(1)
Darknet IPs are not assigned to actual hosts so the traffic only captures the initiation of a communication, and not the actual channel traffic (e.g., payloads). This is in contrast to honeypots which emulate real services (e.g., a Web application or SSH server).
As a result, only metadata such as incoming packet's source IP (src IP), destination IP (dst IP), destination port (dst port), and packet size are available. 
(2) Darknet traffic is full of benign scanning activity from services/enterprises such as Amazon, Google, and Shodan which must be filtered out. 
(3) Attackers repackage old malwares (e.g., Mirai) and reuse known attack patterns which makes it difficult to identify novel threats.
(4) Efficient algorithms are a necessity since terabytes of darknet data is generated every month. 

Recent works proposed various methods for analyzing darknet traffic. 
Since the destination TCP or UDP port number provides a good indication of the sender's intentions (e.g., accessing port 23 may indicate an attempt to search for an accessible Telnet server), 
most of the previous research has focused on grouping ports into static clusters and detecting peaks or unusual trends in the volume of the clusters or individual ports~\cite{Pa2016, ban2012, Fachkha2015, Bou-Harb2015}.
However, attacks are becoming more sophisticated, automated, and noisy (in terms of port access). For example, attackers may perform multi-stage attacks or attempt to exploit multiple vulnerabilities to compromise a device ~\cite{singhal2017security}. 
The methods proposed in previous works (1) cannot identify attacks which use more than one port in sequence, and (2) only perform one time clustering, and therefore do not provide any means for tracking on-going attacks, detecting a recurring/reused attack, or identifying novel emerging threats over time. 

In order to detect attacks it is important for a security analyst to be able to analyze darknet data and provide insights on an hourly basis.
However, this analysis is challenging as there are terabytes of darknet traffic data every month and this figure is expected to increase in the coming years. 
We believe the solution to this challenge will be based on utilizing the power of big data and using a distributed algorithm to provide hourly reports and alerts.

In this paper, we propose \textbf{Da}rk\textbf{n}et \textbf{T}raffic \textbf{E}mbedding (DANTE), a novel darknet analysis method for representing, monitoring, and detecting complex emerging threats via darknet traffic. DANTE is designed to be scalable and work in a big data architecture.
DANTE accomplishes this in two steps. First, for each time window, DANTE summarizes each hosts' activity as a feature vector which captures that host's behavior. The vector is made by (1) training a Word2vec~\cite{mikolov2013} model on the targeted ports (words) found in observed sequences (sentences), and (2) summarizing newly observed port sequences from a host as the average of the ports' Word2Vec embeddings.
The second step is to cluster the feature vectors (host activities) found in a time window, and use a novel technique to map these clusters to the previous time windows. Briefly,  (1) the cluster labels from the previous time window are mapped to the current clusters using Jaccard similarity measures (\textit{tracking}), (2) clusters which could not be mapped are then labeled using a collection of one-vs-all classifiers (\textit{recurring concepts}), and then (3) the remaining unlabeled clusters are labeled by a members of the CERT and are given a new 1-vs-all classifiers (\textit{novelty detection}).

To evaluate our method, we worked with the largest telecommunications provider in Europe, Deutsche Telekom AG.
To support our research, they collected for us full year of darknet traffic (over three Tera-Bytes) in which DANTE discovered 1,177 new emerging threats and was able to track malicious campaigns over time. Deutsche Telekom's is now using DANTE for threat intelligence in their CERT. We also evaluated the current best approach \cite{ban2016} on the same dataset and found that DANTE produced results were more concise and significantly more effective at capturing attack patterns.

In summary, this paper provides the following contributions:
\begin{enumerate}

\item A scalable and online framework for analyzing darknet traffic which can detect and track malicious campaigns and behaviors over time.

\item A method for representing a sequence of accessed ports, of variable length, as a numerical feature vector which captures the intent of an attacking host in a meaningful way.

\item A generic algorithm for performing temporal clustering which can (1) track cluster drift, detect novel/emerging clusters and reoccurring clusters, (2) run parallelized over a big data cluster with multiple data sources, and (3) can be used with any batch clustering algorithm according to the user's needs.

\end{enumerate}

The remainder of the paper is organized as follows. 
Section 2 provides background information on darknet-related research and embedding techniques. 
The proposed darknet monitoring framework is presented in Section 3. 
In Section 4 we present the proposed approach for port scan embedding.
Section 5 contains a description of the novel temporal clustering method. In Section 6 we present our experimental results and in Section 7 we conclude the paper.

\section{\label{sec:relatedWork}Related Work}

The two main contributions of this paper are (1) a method for analyzing darknet traffic, and (2) a method for clustering time series data using static clustering algorithms. 
Therefore, in our review of related work, we address each subject separately.

\subsection{Mining Darknet Traffic}

In prior research ~\cite{bailey2005internet, bailey2006practical, ban2015, ban2012, Ban2017, Bou-Harb2015, Fachkha2015, Furutani2014, Liu2014, Pang2016, Choi2013, skrjanc2017}, darknet data is used to detect botnet hosts, typically by clustering and classifying the src IPs with features such as the dst port and packet size. 

In \cite{Liu2014}, the authors created a rule-based model to help categorize darknet records into several types of malicious attacks and benign activities, and showed how those categories evolved over ten years of data. 
The authors used attributes such as the number of source IPs and destination ports in order to categorize the data. 
However, they did not consider the sequence of destination ports coming from an IP. 
We found those sequences to be particularly informative in the detection of attack patterns as they can indicate the intention of the attacker.

Ban \etal~\cite{ban2012, ban2016, Ban2017, ban2015} introduced a network incident analysis center for tactical emergency response (NICTER) that monitors around 300,000 darknet IPs in Japan. 
They used NICTER to find correlations between the malicious activities discovered on the darknet and activities extracted from different types of honeypots. 
In~\cite{ban2015}, and later in~\cite{ban2016}, Ban \etal used DT-growth, an association rule learning (ARL) algorithm, to port associations in the darknet data. 
Their research serves as the foundation of this paper, as they showed that many attacks patterns use more than one port and thus should be grouped. 

Thonnard and Dacier~\cite{thonnard2008} created a new clustering tool to detect groups of IPs that behave similarly. 
They used graph theory in order to find temporal correlation between port usage and thus created a way to group different IPs. 
Nevertheless, this work is different from ours, as Thonnard and Dacier~\cite{thonnard2008} ignored the meaning and use of the ports in the sequence when clustering.

In~\cite{coudriau2016topological}, the authors used DBSCAN to create clusters of packets and then used an algorithm from the field of topological data analysis in order to visualize the darknet and help an expert easily observe and analyze the data. 
To use DBSCAN, they treated the ports as integers by looking at the port number. 
In contrast to this work, we use an artificial neural network to learn the connections and relations between the ports to find an informative numeric representation.

In order to retrieve numeric information from network traffic packets, the majority of the research conducted extracted statistical features such as the number of destination IPs or unique ports~\cite{Pang2016, Lagraa2017, Owezarski2015,  Casas2012, Corchado2010, Bartos2016, Zhang2016, coudriau2016topological}. 
Although these features proved to help in the detection and exploration of attacks, they are hand-picked, and it is difficult to choose the features that fit the task. 
In contrast to those methods, we used a neural network-based algorithm that automatically extracts meaningful representations of the packets.

Most of the aforementioned works apply their method on a static corpus of data.
However, new data arrives continuously, and there is a need for an online system that can detect attacks in near real-time. 
To address this issue, our proposed method periodically analyzes the packets that have arrived from the sensor in the last $L$ minutes and applies the detection mechanism in an online fashion by using a big data architectures.

\subsection{Temporal Clustering}

The well-known of clustering algorithms, such as k-means and DBSCAN~\cite{ester1996density}, are batch algorithms. 
This means that they are applied once on the entire dataset and cannot track or monitor temporal trends. 
Although batch algorithms provide the best clustering quality,  they are unsuitable as-is for processing data streams (unbounded sequences of observations).

To cluster data streams, one can use STREAM~\cite{guha2000}, Incremental DBSCAN~\cite{ester1998}, DenStream~\cite{cao2006}, CluStream, and many others~\cite{Carnein2019}.
However, these algorithms cannot perform novelty detection. 
In other words, they cannot differentiate between reoccurring clusters and novel clusters. 
This capability is required to detect emerging threats and monitor the re-use of known attack variants.

There are several algorithms for novelty detection in data streams  \cite{faria2016novelty}, however these algorithms cannot be natively parallelized over a big data computing cluster and do not directly support the processing of multiple parallel sources. 
Furthermore, each of these algorithms were designed to apply the principles of a particular kind of clustering algorithm over compressed data summaries. 
In addition to this being a lossy process, a user may need a different type of clustering algorithm to best fit the data (e.g., DBSCAN or K-means)).
In contrast, the temporal clustering framework that we propose in this paper can be parallelized over a big data cluster while receiving data from multiple sources.
In addition, the framework is flexible in terms of selecting a clustering algorithm. 
This enables the user to apply the most suitable batch algorithm in his/her arsenal.

\begin{figure*}[t]
\centering
\includegraphics[width=1\linewidth]{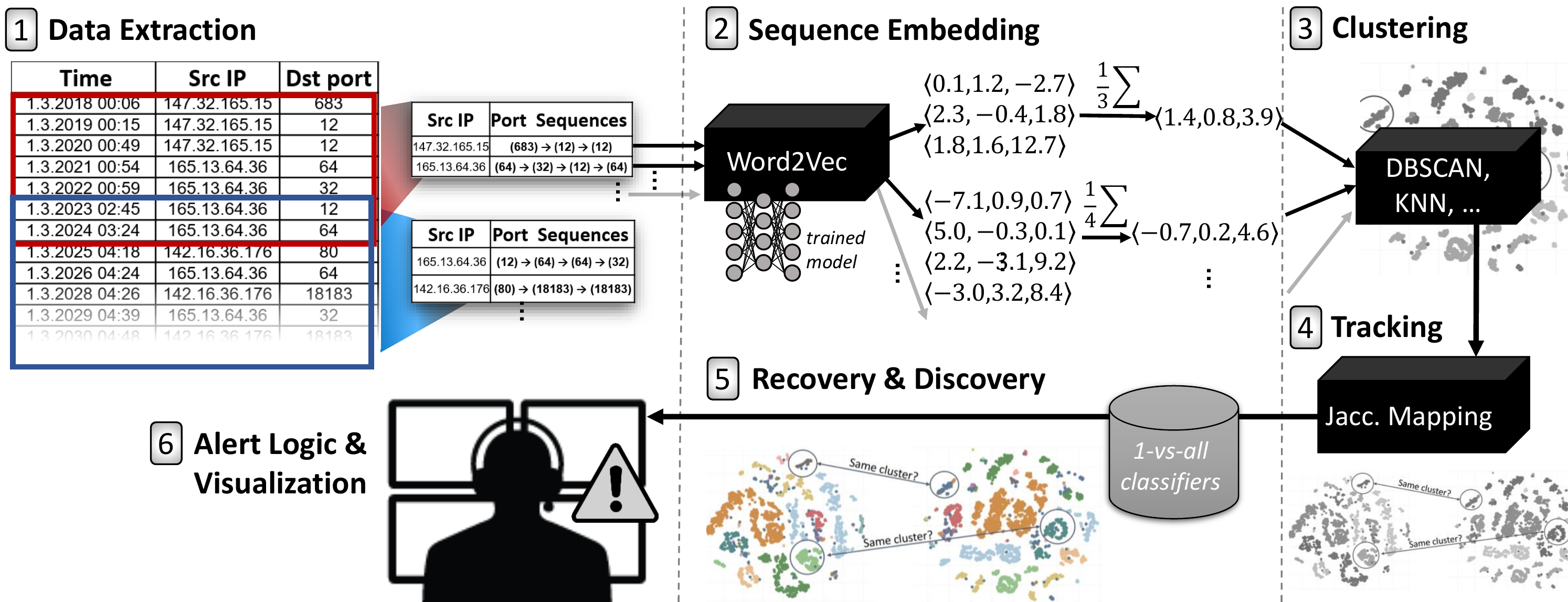}
\caption{An overview of the DANTE framework}
\label{fig:summerized_flowv}
\end{figure*}


\begin{figure}
\centering
\includegraphics[width=1\linewidth]{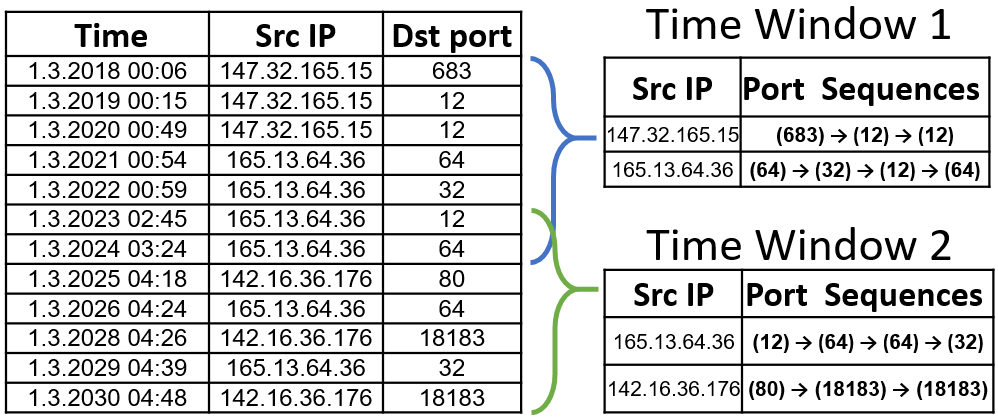}
\caption{Time window extraction.}
\label{fig:events_extraction}
\end{figure}

\section{\label{sec:framework}The Darknet Analyzer Framework}
In this study, we analyze the ongoing activities in the darknet by clustering common patterns/sequences of targeted (dst) ports.
These patterns can be used to discover new attacks as well as explore the behavior of ongoing attacks and trends.

The analysis process consists of the four stages described below.

\begin{enumerate}
\item \textbf{Sequence extraction}
First, the data is split into sliding time windows, resulting in multiple windows with length L; for each time window, we group the destination port records of the same src IP into a port sequence. 
The final result is a table representing a time window with two attributes, the first being the src IP, and the second being the dst port's sequence, as shown in Figure~\ref{fig:events_extraction}.

\item \textbf{Port sequence embeddings} 
By using a word embedding algorithm on the port sequences extracted from the previous stage and treating ports as words and port sequence as sentences, we are able to transform the port sequences into a meaningful numerical feature vectors. 
A description of how we expand on this method is presented in Section~\ref{sec:PortSequenceEmbbedings}.

\item \textbf{Temporal clustering} With the feature vector obtained in the previous stage, we use a novel temporal clustering method to cluster the feature vectors over time. 
This step enables us to track attack patterns and distinguish between new (novel) and reoccurring patterns.
This is explained further in chapter \ref{sec:TemporalClustering}.

\item \textbf{Alert logic and visualization} 
The current time window's clusters and labels are visualized for the CERT team (e.g., trends, or campaigns by country source).
We can also use this information to create an alert rule, e.g., an alert regarding the reappearance of a cluster that an analyst classified as malicious in the past. 
These alerts can then be used in security information sharing platforms such as MISP~\cite{misp2016}.
\end{enumerate}

In addition, the above-mentioned alert system will allow us to use DANTE to deal with adversarial attacks. 
Such attacks can be divided into two groups. In the first, the attacker is trying to conceal him/herself by adding dummy port access as noise. 
A simple way to deal with this attack group is to include an alert rule that issues alert when a never seen before cluster is seen, as those attacks will create a new cluster. To avoid a mass amount of false positive, those alarm are only sent if a cluster is larger than a certain threshold. 
In the second attack group, the attacker will try to disguise him/herself as a pattern that belongs to a known cluster, such as a cluster that consists of a popular port sequence pattern. 
To deal with this group, one can create an alert rule to issue an alert when a cluster dramatically increases in size. 
Another way to deal with this group of attacks is to recluster the large clusters and find subpatterns within them; those clusters could help an analyst find malicious subpatterns that differ from the other subclusters and those can indicate hidden attempts.

\section{\label{sec:PortSequenceEmbbedings}Port Sequence Embedding}

Threat agents (e.g., an attacker or bot) may send packets to unregistered IP addresses for several different reasons, such as to find a host with a vulnerability to exploit
or, in the case of worms, to access a backdoor.

We define a sequence $S$ as the sequence of ports collected from a specific src IP to a specific dst IP.

A darknet sensor can observe these communications as a sequence of ports being accessed. 
For example, the sequence ``42527, 80, 80'' was observed in the wild. 
In this sequence, we can see that the attacker tried to access a high port (42527), and immediately send several packets to port 80 (HTTP); This sequence reflects a worm's attempt to detect if the target host has already been infected (via backdoor on port 42527), and then after failing, trying to compromise the host by exploiting two different vulnerabilities in the host's web server (port 80).
From this, we can understand that the port targets in a given sequence reveal information regarding the intent of the attacker. 
From this example, it can be seen that the sequence of targeted ports not only reflects the attacker's goal and strategy, but also implicitly captures the type of threat agent. Moreover, threat agents such as bots, worms, and hackers in the same campaign act in unison enabling us to perform cluster analysis to detect new or recurring campaigns and other behaviors.

However, in order to cluster those sequences, we must find a representation which can summarize them as a numeric vector for the machine learning algorithm. This format is a fixed length vector $x \in \mathbb{R}^n$ where the euclidean distance between vectors $x_i$ and $x_j$ vectors measure similarity (closer vectors are more similar). 

Although TCP and UDP ports are numbers, the numerical relationship between ports is meaningless. 
For instance, port 21 is used for FTP, and port 22 is used for SSH, and there is no connection between the two. 
Therefore, in order to summarize the behavior of a scan, we first need to learn a numeric relationship between all of the ports. 

To accomplish this we use Word2Vec.
Word2vec, presented in~\cite{mikolov2013} by Mikolov \etal, is a natural language processing (NLP) algorithm that aims to maximize the co-occurrence probability of words in the same sentence. Our method uses the same basic algorithm, but instead of looking at words in sentences, we use the port sequences where a sequence from a specific src IP corresponds to a sentence, and the port numbers correspond to the words in that sentence.
Let $p$ be a TCP/UDP port in the set $P=\{0,1,\ldots 65,536\}$ of all ports. Let $s = \{p_1,p_2, \ldots p_k\}$ be a sequence of ports sent from a specific src IP to a specific dst IP within some time interval $L$. We train a Word2Vec model on a corpus of example sequences $S$, such that a sequence $s \in S$ is treated as a sentence and $p \in P$ is treated as a word. 

The trained Word2Vec model is able to convert a port $p$ into a vector representation (embedding) $e_p \in \mathbb{R}^n$ such that $e_p$ captures the meaning of $p$ given its observed locality among other ports in sequences of $S$.
An example of that property can be seen by looking at port 23 and port 2323, both of which are used for Telnet and hence are expected to appear in the scan data interchangeably.
Therefore, they have remarkably similar embedding vectors. The same applies to arbitrary port numbers which are dynamic or not well documented.
By using Word2vec, we do not need to consider the fact that multiple ports use the same service as the embedding process does this for us. 
Moreover, ports commonly found together in particular attack patterns will be associated as well, thus $e_p$ captures the attack intent as well. 

Unlike more straightforward methods, such as Bag of Words (BoW), Word2vec uses context windows in order to create the representation of each word. This method considers not only what other words are in the sentence, but also where they are, allowing for a better representation.

In order to build this port-to-embedding transformation model we need to supply it with a significant amount of scan data, which could be computationally heavy. Fortunately, this model does not have to be rebuilt in every time window, and it is possible to use a pretrained model for a long period of time. The intuition behind this rationale is that the uses of each port do not change often and a well-trained model should be sufficient for a considerable amount of time. 

Additionally, there is no need to save the model itself once trained. Instead of keeping the entire neural network model, one can save a hash table where the key is the port number, and the value is the embedding for that port. This approach reduces the amount of data needed to be saved significantly, as the number of possible ports is limited by the number 65,536. Further more, by eliminating the need to execute a neural network, DANTE can run extremely fast in a big data framework.

After each port has an embedding vector of size \(d\), we want to obtain an embedding vector with the same size, \(d\), that represents an entire port sequence \(P\) that contains $s$ number of ports. 
Although there are many methods for sentence embedding, recent research~\cite{wieting2015towards} discovered that the best way to do so is to average the embedding of each word in the sentence. In the port embeddings case, We summarize the overall behavior (intent) of $s$ as a single vector $e_s \in \mathbb{R}$ by computing the average of that sequence's port embeddings. Concretely, we compute

\begin{equation}\label{eq:embed}
e_s = \frac{1}{k}\sum_{i=1}^{k}{e_{p_i}}
\end{equation}

The resulting feature vector can be used for any machine learning algorithm, such as a classifier or clustering algorithm.

\section{Temporal Clustering}
\label{sec:TemporalClustering}

As described in Section~\ref{sec:PortSequenceEmbbedings}, it is possible to summarize sequences of ports as their average embedding and analyze their behavior by performing cluster analysis. 
However, it is important to inspect the clusters over time. 
By doing so we can: (1) detect new attacks as they emerge (novelty detection), (2) track attack campaigns and how their strategies change, (3) follow the re-use of known attacks, e.g., variants of the Mirai botnet, and (4) analyze the trend of ongoing attacks, such as changes in volume, sources, and targets.

However, darknet data is collected from $X$ sources simultaneously. 
To manage this, the data it is typically stored in a big data cluster such as Hadoop. 
Therefore, we propose a temporal clustering framework which can be used with any batch clustering algorithm. 
The framework operates as follows.

\subsection{Windowing}
 First, we sort and aggregate the most recent data into overlapping time windows. 
 Let $L$ be the width of the window in minutes, and let $S$ be the step size in which we slide the window, where $S<L$. 
 Following this process, let $T_{i}$ be the $i$-th time window in our data, where $T_{i+1}$ is the next sequential time window.
 Finally, let the ratio of observations shared between two neighboring windows be defined as
\begin{equation}
    r_{i,i+1} \frac{|T_{i}| \cap |T_{i+1}|}{L}
\end{equation}
The overlap between neighboring windows is necessary in order to track clusters. 
To ensure this, the parameter $S$ should be small enough so that
$0.2 \le  r_{i,i+1} \le 0.8$.

\subsection{Clustering}
Next, we apply a clustering algorithm to the data of each time window to group the observations. 
We note that any batch clustering algorithm can be used. 
For example, K-means, Fuzzy C-means, Gaussian mixture models, hierarchical clustering, spectral clustering, and more. 
For our dataset, we found that the clustering algorithm, DBSCAN~\cite{ester1998}, worked best. 
The reason is because DBSCAN clusters data based on density. 
As a result, the number of clusters discovered is variable and does not need to be predefined (as in k-means). 
Another advantage is that DBSCAN can label outliers (points which are relatively far from the general distribution). 
This helps us analyze these cases separately without harming the quality of the clustering process.

\subsection{Tracking}
\begin{figure}[t]
\centering
\includegraphics[width=1\linewidth]{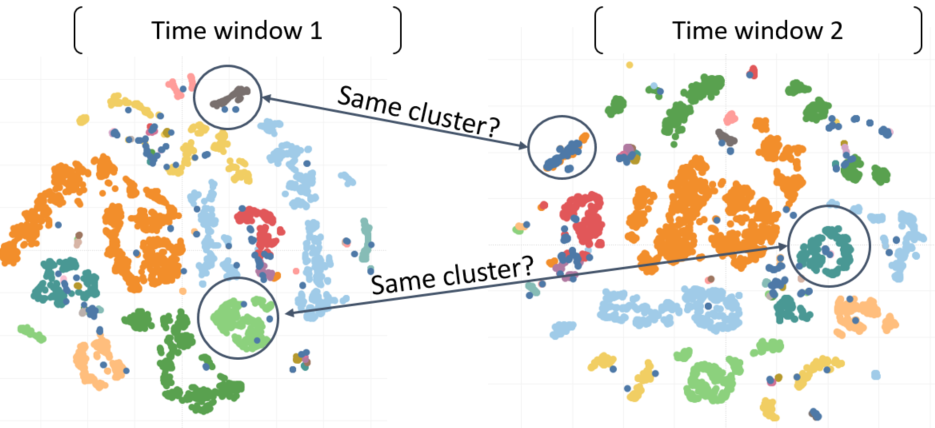}
\caption{An illustration of the challenge in mapping clusters found in different time windows (data here has been projected as 2D using T-SNE~\cite{maaten2008visualizing}).}
\label{fig:clutser_tracking}
\end{figure}

Between time window $T_{i}$ and time window $T_{i+1}$, concept drift can occur. This means that the number of clusters and their shapes can change.
Moreover, a cluster in $T_{i+1}$ can be a $current$ cluster (also found in $T_{i}$), an $old$ cluster (found in $T_{j}$ where $j<i$), or a $new$ cluster (never seen before).
Figure~\ref{fig:clutser_tracking} illustrates this challenge.

To annotate the clusters in $T_{i+1}$, we first find the current clusters by comparing $T_{i}$ and $T_{i+1}$. 
A cluster in $T_{i+1}$ is mapped to a cluster in $T_{i}$ if there is a significant overlap of observations between them. 
We measure the overlap using the Jaccard similarity metric, defined as
\begin{equation}
    \text{Jaccard}(A,B) = \frac{|A \cap B|}{|A \cup  B|} = \frac{|A \cap B|}{|A| + |B| - |A \cap  B|}
\end{equation}

The Jaccard similarity metric measures the similarity between sets of items. 
This metric can be used in our case, because adjacent time windows overlap (by $L-S$). 
As a result, clusters which have a high Jaccard Similarity Score have a large number of overlapping observations and thus are considered to be the same pattern.

By using the distributed system, we simultaneously calculate the Jaccard similarity of all of the clusters in $T_{i+1}$ with the clusters in $T_{i}$. 
If the Jaccard similarity is above a certain threshold for two clusters, then the cluster from $T_{i+1}$ is considered to be the same as the cluster from $T_{i}$ (i.e., $current$ cluster). 
In cases in which the $T_{i+1}$ cluster has no corresponding cluster from $T_{i}$, the cluster is considered new. 
The algorithm for mapping clusters between adjacent overlapping time windows is presented in Algorithm~\ref{clus2clus}. 
Note that there is no need to use the embedding vector of each instance; only a key (the src IP in our case) is needed for the comparison. 

\begin{algorithm}[h]
\begin{algorithmic}[1]

\Require $T_i,T_{i+1},Threshold$
\Ensure $ClusterToClusterMapping$

\State{$ClusterToClusterMapping \gets emptylist \;$}
\ForAll {$p \in T_i $}:
    \ForAll {$k \in T_{i+1} $}:
        \State{$Sim \gets Jaccard(p,k)$ \;}
        \If{$Sim > Threshold$}{
            \State{ $ClusterToClusterMapping[p] \gets k$ };
        \EndIf
        }
    \EndFor

\EndFor
\end{algorithmic}

\caption{The cluster mapping algorithm for the current and previous time windows\label{clus2clus}} 

\end{algorithm}

\subsection{Recovery \& Discovery}

\begin{algorithm}[h]
\begin{algorithmic}[1]
\Require{$T_{I,c}, MS, Threshold$}
\Ensure{$TrackedCluster$ }


\State $N\gets NumberOfInstances(T_{I,c})$
\State $TrackedCluster\gets emptyString$ 
\ForAll {$M \in  MS $}:
    \State $C \gets 0$ 
    \ForAll {$I \in  T_{I,c} $}:
        \State{$P_I \gets P(Y_i = 1 | M)$}
        \State{$C \gets C + P_I$}
    \EndFor
    \State{$C_{normalized} \gets C / N$}
    \If{$C_{normalized} > Threshold$}{
        \State {$TrackedCluster\gets M.name$}
    }\EndIf
\EndFor

\end{algorithmic}
\caption{The recovery \& discovery algorithm for finding older appearances of a cluster  \label{past_models} }
\end{algorithm}

\begin{figure*}[t]
\centering
\includegraphics[width=1\linewidth]{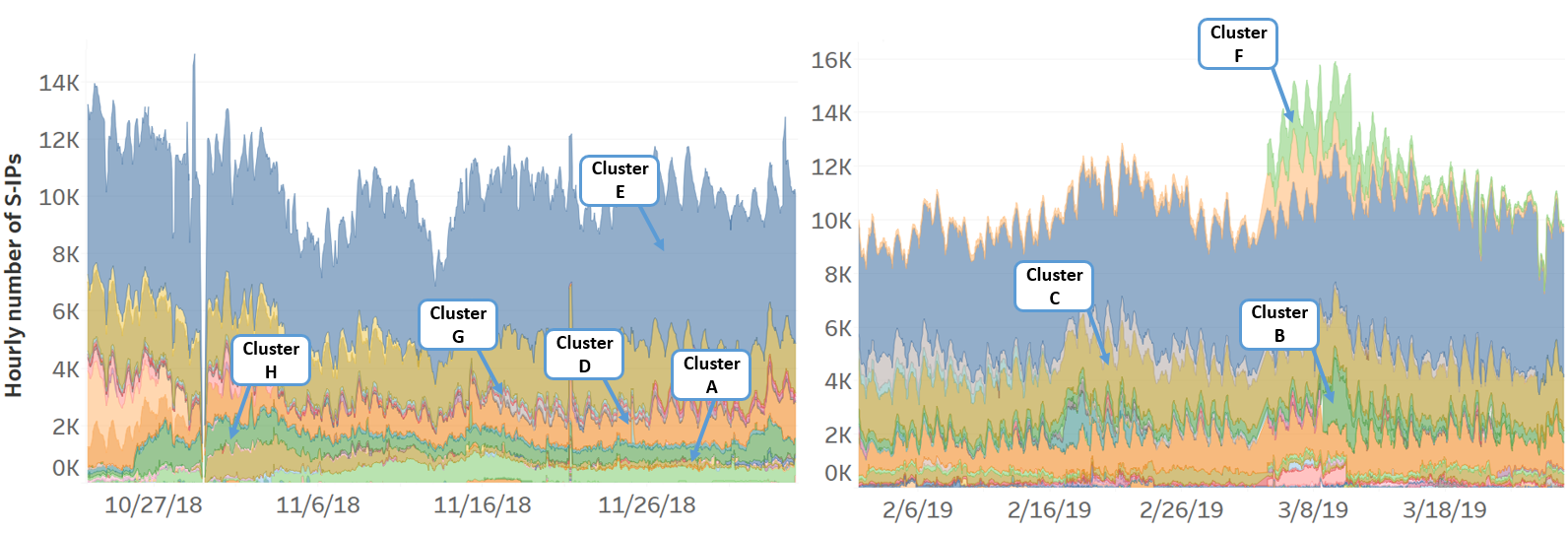}
\caption{An overview of the different clusters in the data during the data collection period, aggregated by hour. Each color represents a different cluster.}
\label{fig:overview}
\end{figure*}

The cluster mapping process presented in the previous Section enables us to align the clusters with the previous time window, but we also want to be able to identify witch of the clusters are recurring concepts (to retrieve
their annotations), as well as novel concepts.
Because storing the entire past data is, in most cases, impractical, our approach is to build a classifier model for each of the observed clusters.
Each model is a binary one-vs-all classifier trained on the time window where the said cluster was first seen. 
The instances that belong to the cluster get the label one, and the rest get the label zero. 
We found that Random Forest suits this problem well as this model, unlike classifiers such as K nearest neighbors, have no need to save the data points and only need to save the decision trees.
We define the set of one-vs-all classifiers as $M$.

Let $M$ be our database of our one-vs-all classifiers and let $c$ be a cluster in $T_{i+1}$ which we could not map to $T_{i}$. The probability that $c$ belongs cluster $c_j$ represented by $m_j \in M$, is computed as
\begin{equation}
	p(c, c_j)=\frac{1}{|c|}\sum_{i=1}^{|c|} \text{predict}_{m_j}(e_{s_i})
\end{equation}
where $\text{predict}_m(x)$ returns the probability that $x$ belongs to the positive class using model $m$. We assign $c$ with the annotation of cluster $c_j$ if $p(c, c_j)$ obtains the highest probability for all models in $M$, and $p(c, c_j) > \beta$ for some user defined parameter $\beta$ (we set $\beta=0.7$). 
Similarly to the Jaccard similarity calculation, one can easily distribute the prediction part as those predictions can also be calculated simultaneously.
A formal description is described in Algorithm~\ref{past_models}.

In cases in which there is no match in any of the classifiers in $M$, we consider cluster $c$ to be a $new$ cluster. Once a new cluster is found we train a new classifier on this cluster's data as previously explained.

After some time, a concept drift may occur, and the patterns change slightly. 
To deal with this issue, in cases in which a known cluster appears in the data stream, we update and retrain the corresponding model.

\section{Analysis of Darknet Traffic}
Unlike other methods that used darknet data streams, e.g. \cite{Choi2013}, DANTE is not trying to find anomalies on specific ports, but rather find concepts and trends in the data. 
While methods that find correlations between ports exist \cite{ban2016}, (1) they operate offline detecting patterns months after the fact, and (2) do not track the patterns over time. In contrast, DANTE finds new trends online (within minutes) and can detect recurring and novel patterns.
To the best of our knowledge there are no online algorithms for detecting and tracking patterns in darknet traffic.
Therefore, we demonstrate the usefulness of the proposed approach through a case study on over a year of data.
The data was collected from a greynet~\cite{harrop2005greynet, ban2012}, meaning that the unused IPs are from a network that is populated by both active and unused IP addresses.

\subsection{Configuration and Setup}
\begin{figure*}[t]
\centering
\includegraphics[width=1\linewidth]{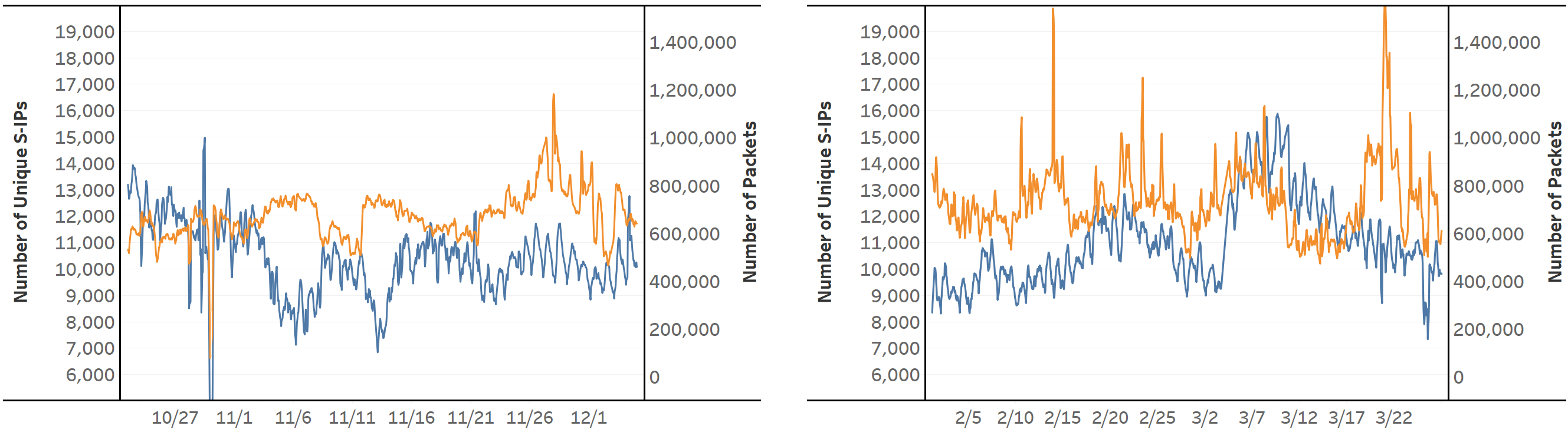}
\caption{The hourly number of different packets (orange) unique src IPs (blue) in the darknet after the data preprocessing stage.}
\label{fig:data_counts}
\end{figure*}

\paragraph{Dataset.} 
For the purpose of this research, the NSP established 1,126 different unused IP addresses across 12 different subnets. 
All traffic which was sent to these IPs were logged as darknet traffic. The traffic was collected in three batches; the first was recorded during a period of six weeks (44 days) from 10/25/2018 until 12/5/2018 (denoted by Batch 1), the second was recorded during a period of eight weeks (55 days) from 2/1/2019 until 3/26/2019 (denoted by Batch 2) and the third batch was recorded during a period of 37 weeks (257 days) from 21/1/2019 until 10/6/2019 (denoted by Batch 3). Note that Batch 2 and Batch 3 have 36 days of overlapping, that is because the Batches 1 and 2 recorded in the research phase of the project, where Batch 3 recorded in the deployment stage, in real time.
A deep analysis was performed on Batches 1 and 2. Batch 3 was used to demonstrate the system's long-term stability.
In total, 7,918,787,884 packet headers from 4,887,568 different source IP addresses were recorded, resulting in over 3 terabytes of data.
Figure~\ref{fig:data_counts} shows the number of packets and source IP addresses for every hour in the first two batches. 
Note that due to a technical problem, one hour at the end of October is missing.
Because the missing time is insignificant, and the proposed method can deal with missing time windows, those missing values do not affect the overall results. 

\paragraph{Configuration.} We choose the step size $S$ to be one hour and the window length $L$ to be four hours, similar to the work of Ban \etal~\cite{ban2016}.
A one hour step size provides a sufficient amount of data while granting a security expert enough time to react to a detected attack. 
In addition, we choose the epsilon parameter of DB-SCAN to be 0.3, and the minPts parameter to be 30, as those parameters resulted in an average of four new clusters every day (agreed by the security experts to be a reasonable number of clusters to investigate each day). 
In addition, we do not want clusters with a small number of src IPs, as those clusters are too small to represent a significant trend in the network, and thus should be treated as noise. 

\paragraph{Scalable Implementation.} To scale to the number of sources and amount of data, DANTE was implemented and evaluated at the NSP's CERT using Spark a Hadoop architecture. 
We tested the method on a Hadoop cluster consisting of 50 cores and 10 executors. 
The algorithm takes approximately 62 seconds to extract, embed, cluster, and map each four hours time window.



\begin{figure}[t]
\centering
\includegraphics[width=1\linewidth]{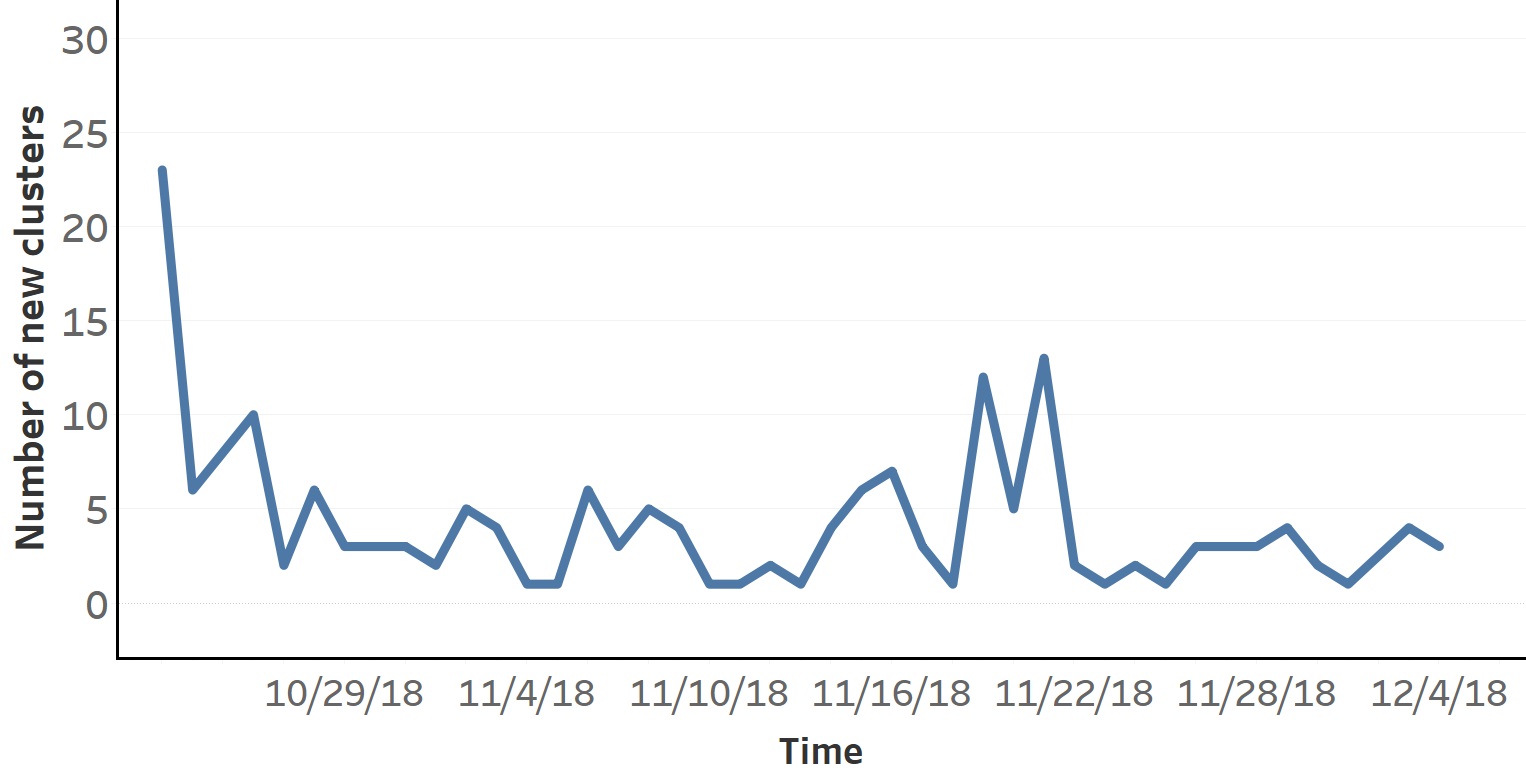}
\caption{Number of newly discovered clusters over time in the Batch 1 data (daily).}
\label{fig:new_clusters_count}
\end{figure}

\paragraph{Data Preprocessing.} As most of the source IPs in the dark data only sent one or two packets during the period of data collection, we decided to remove them as those port sequences associated with those IPs are too short and cannot constitute a meaningful pattern. 
Filtering those IPs reduce the noise and therefore improve the results. 
In addition, some of those IPs are likely to be a random miss configuration and not an active malicious attack. 
By removing those IPs, we reduced the number of packets by 33\% percent.

\subsection{Results}
\begin{figure}[t]
\centering
\includegraphics[width=1\linewidth]{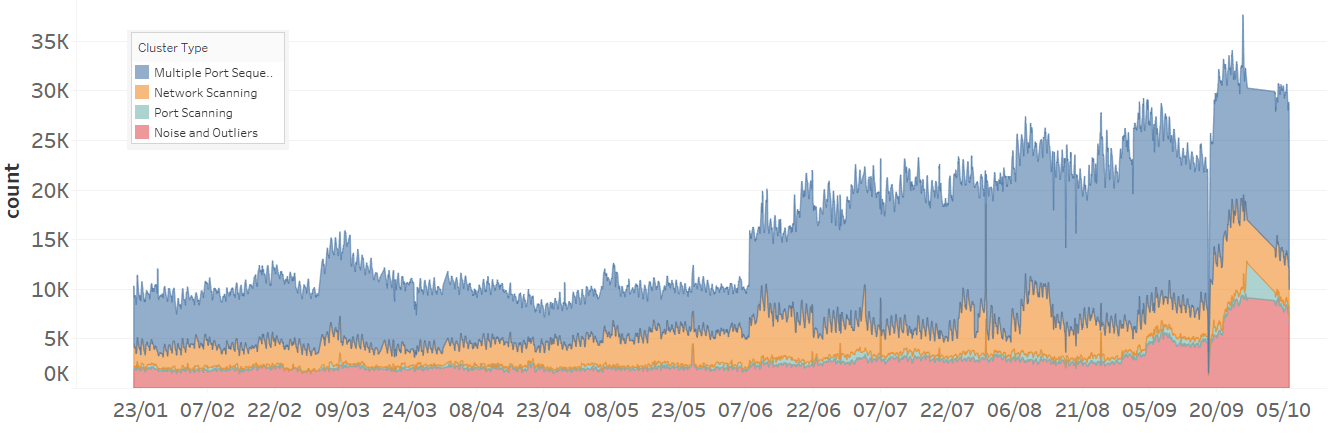}
\caption{Hourly number of sequences over 10 month of data, by their types.}
\label{fig:year_by_types}
\end{figure}

Because of time constrains, we use Batch 1 and 2 for a comprehensive analysis, and show that DANTE can find malicious attack within hours, long before the first online report. 
Batch 3 is used to prove that the algorithm can work over a long period of time, and can detect new and reoccurring clusters even after almost a year. 
A visualisation of Batches 3 clusters over the 10 months can be seen in Figure~\ref{fig:year_by_types}.

A total of 400 clusters (novel darknet behaviors) were discovered over Batches 1 and 2, and 1,141 clusters were discovered over Batch 3. 
As previously mentioned, the system discovered four new clusters each day, on average, as can be seen in Figure~\ref{fig:new_clusters_count}. This number, determined by the set parameters, is small enough for an analyst to evaluate but large enough to detect new concepts. 
We found that the vast majority of the src IPs belong to one of 16 large clusters (concepts) over the year. Some of these clusters reoccurred (e.g., botnet code reuse) and others always stayed present (e.g., worms looking for open telnet ports).
Figure~\ref{fig:overview} plots the volume of the discovered clusters over time for Batches 1 and 2.
The $x$-axis represents time, and the $y$-axis represents the total number of src IPs in the data for each hour. 
In the visualization component of DANTE (section \ref{sec:framework}) this graph is updated periodically to help the analysts explore past and current trends. 

The clusters discovered by DANTE can be roughly divided into four categories:

\begin{enumerate}
    \item \textbf{Service Recon} Clusters whose sequences have six or more unique ports:\textit{Either benign web scanners or agents performing reconnaissance on active services. }
	\item \textbf{Basic Attacks \& Host Recon} Clusters whose sequences consist of a single port: \textit{Agents trying compromise a device via a single vulnerability, or performing reconnaissance to discover active network hosts.}
    \item \textbf{Complex Attacks} Clusters whose sequences have two to five unique ports: \textit{Agents which are attempting to exploit multiple vulnerabilities per device, are performing multi-step attacks, or are worms.}
    \item \textbf{Noise and Outliers} A single cluster of benign sequences from misconfigurations, backscatter, or are too small to represent an ongoing trend.
\end{enumerate}

In Table~\ref{cluster_families} we give statistics on these categories, and in Figure~\ref{fig:year_by_types} we present the volume of each category in Batch 3 over time. 
Table \ref{cluster_examples} provides concrete examples of discovered attack patterns, by cluster, as labeled in Figure~\ref{fig:overview}.


\begin{table*}[]
\caption{The number of clusters, src IPs, and packets for each of the four cluster families. \label{cluster_families}}

\resizebox{\linewidth}{!}{%

\begin{tabular}{|c|c|c|c|c|c|c|c|c|c|}
\hline
                                 & \multicolumn{3}{c|}{ No. of clusters}                                 & \multicolumn{3}{c|}{ No. of src IPs}                                                 & \multicolumn{3}{c|}{ No. of  packets}                                                          \\ \cline{2-10} 
\multirow{-2}{*}{ Clusters Type} & Batch 1                      & Batch 2                       & Batch 3                       & Batch 1                           & Batch 2                           & Batch 3                             & Batch 1                              & Batch 2                              & Batch 3                                 \\ \hline
Service Recon                                           & 24                           & 27                            & 326                           & 16,909                            & 78,644                            & 101,316                             & 9,603,305                            & 39,111,889                           & 162,499,127                             \\ \hline
Basic Attacks \& Host Recon                                        & 77                           & 71                            & 202                           & 246,864                           & 313,457                           & 1,216,077                           & 50,937,141                           & 56,881,565                           & 380,244,017                             \\ \hline
\textbf{Complex Attacks}        & \textbf{78} & \textbf{121} & \textbf{610} & \textbf{351,318} & \textbf{395,109} & \textbf{2,602,184} & \textbf{19,203,613} & \textbf{85,960,212} & \textbf{1,085,777,577} \\ \hline
Noise and Outliers                                      & 1                            & 1                             & 1                             & 115,645                           & 142,920                           & 819,566                             & 582,646,403                          & 743,977,834                          & 5,935,419,695                           \\ \hline
\end{tabular}}
\end{table*}

\newcounter{ClusterExample}
\addtocounter{ClusterExample}{1}

\vspace{5mm}

\subsubsection{Service Recon}
Service reconnaissance (port scanning) is performed by attackers to find and exploit backdoors or vulnerabilities in services ~\cite{Liu2014}.
In Batches 1 and 2, we identified 51 port scanning clusters with an average of 929 different ports scanned in each cluster.
However, it is important to note that a cluster in this group is not necessarily malicious. 
For example, cluster $\Alph{ClusterExample}$ in Figure~\ref{fig:overview} consists of src IPs of Censys, a security company which scans 40 ports to find and report vulnerable IoT devices. Interestingly, because of the embeddings, DANTE found that $2.2\%$ of the IPs in cluster $\Alph{ClusterExample}$ do not belong Censys' subnet. Rather, we verified them to belong to a malicious actor copying Censys to stay under the radar.

\stepcounter{ClusterExample}

In some cases, the service recon can consist of multiple ports that belong to the same service, in order to use an exploit on this service even if the host is using an alternative port. 
For example, Cluster $\Alph{ClusterExample}$, occurred on 3/8/2019, consist only of ports that can be associated with HTTP.
This cluster consists of 17 ports, such as 80, 8080, 8000, 8008, 8081 and 8181.  
In addition, most of the src IPs are located in Taiwan (18\%), Iran (15\%) and Vietnam (12\%). 
Because DANTE assigns similar embedding to those ports, those port were group together and DANTE was able to detect this pattern and issue an alert. 
At the time of writing of this article, we were not able to find any information on this scan online. This lack of reports could be explained by the fact that there was no significant peak in any of the ports involved, which make it hard for conventional anomaly detectors to detect this pattern. 
\stepcounter{ClusterExample}

\begin{figure*}[t]
\centering
\includegraphics[width=1\linewidth]{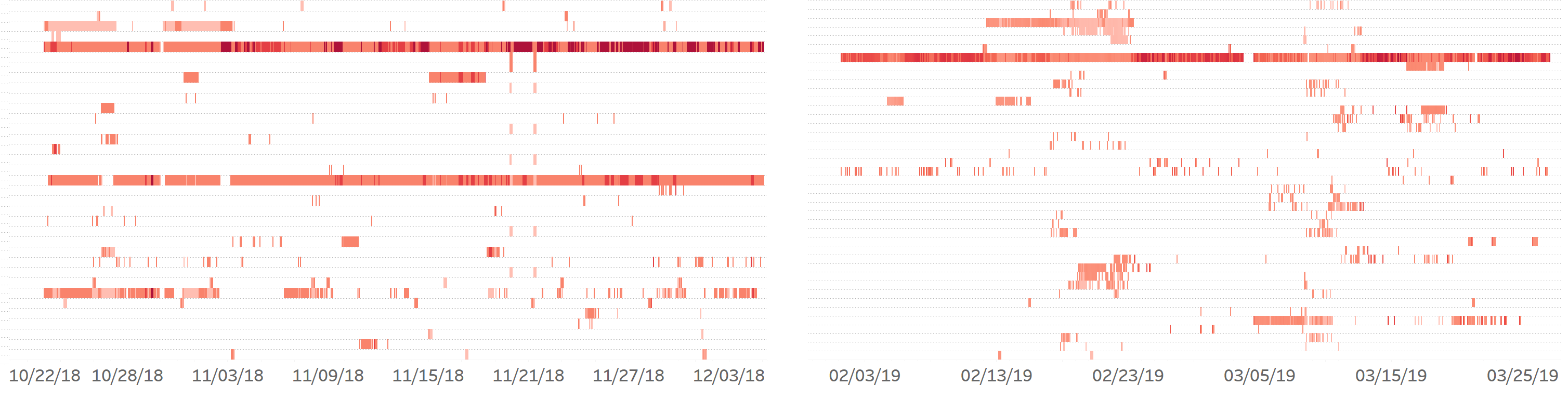}
\caption{
A Gantt chart of cluster re-occurrence. 
Each horizontal bar represents a different cluster. 
Darker color represents more packets in the cluster.
}
\label{fig:multiple_ports_gantt}
\end{figure*}

\subsubsection{Basic Attacks \& Host Recon}

\begin{figure}[h]
\centering
\includegraphics[width=1\linewidth]{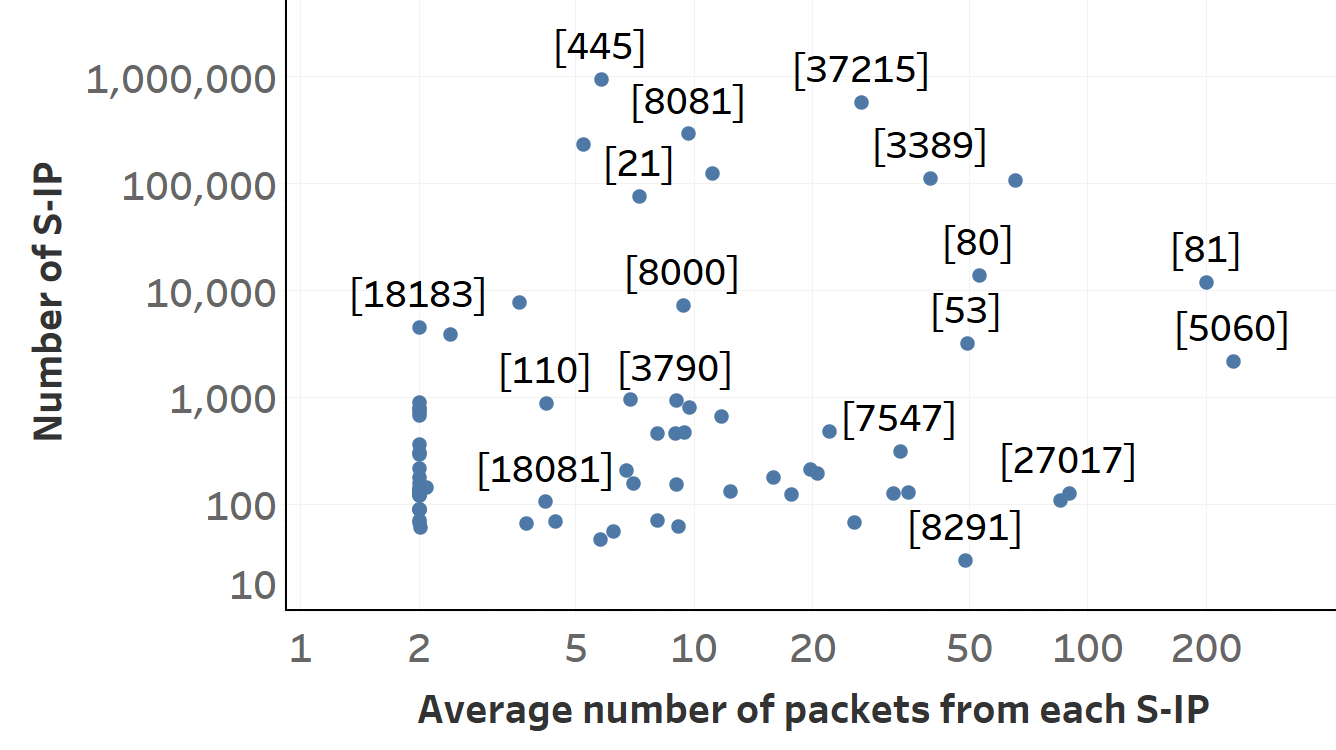}
\caption{
A view of the Basic Attacks clusters from Batch 1. 
Each dot represents a cluster with the corresponding port shown above. 
Note that both the $x$-axis and the $y$-axis are logarithmic.} 
\label{fig:single_port_scatter_plot}
\end{figure}

Many port sequences consist of a single port accessed repeatedly on or more times. This behavior captures agents which are trying to exploit a single vulnerability (e.g., how the Mirai malware compromises devices via telnet).  This behavior is also consistent with network Scanning~\cite{Liu2014}, where the attacker tries to detect live hosts to map out an organization. By following how campaigns (clusters) are reoccur and grow in volume over time, analysts can use DANTE to find trends and discover new vulnerabilities. 
For instance, Figure~\ref{fig:single_port_scatter_plot} contains a scatter plot of each port sequence cluster from Batch 1 and its corresponding port. 


It is interesting to see that, for example, port 445 is accessed by many src IPs, each of which only accessed the port an average of 5.8 times. 
This type of cluster can be seen in the second largest cluster, $\Alph{ClusterExample}$.
This cluster consists of port 445 (SMB over IP: commonly used by worms such as Conficker ~\cite{Durumeric2014, Wustrow2010, heo2018} to compromise Windows machines) and was being attacked at a low rate, once every few hours.
\stepcounter{ClusterExample}
However, DANTE also detected that port 5060 (SIP: used to initiate and maintain voice/video calls) was trending, being attacked at a high rate from the USA.

Nonetheless, most of the clusters in this category are relatively small and only appear for a few hours.
In another example, DANTE detected a network recon campaign between 1:00 and 6:00 AM GMT on 11/25/2018 (cluster $\Alph{ClusterExample}$) on the unused port of 11390.
The attack consisted of 895 different src IPs which mostly targeted one out of our 1,129 darknet IPs. This indicated which subnet the campaign was targeting and provided us with threat intelligence on a possible software vulnerability/backdoor on port 11390.
By using the distributed system, DANTE was able to issue an alert with that information about a minute after the data arrived.\stepcounter{ClusterExample}

\begin{table*}[htbp]
\caption{The number of src IPs, packets, and a sequence example for each of the mentioned clusters. \label{cluster_examples}}
\resizebox{\linewidth}{!}{

\begin{tabular}{|c|c|c|c|c|}

\hline
\textbf{\begin{tabular}[c]{@{}c@{}}Cluster\\   Name\end{tabular}} & \textbf{Category}                                                     & \textbf{Number of src IPs} & \textbf{Number of Packets} & \textbf{Sequence Examples}                                                        \\ \hline
A                                                                 & Service Recon                                                       & 141                      & 873,458                    & {[}2077, 2077, 8877, 7080, \dotso , 9304, 3556{]}                                \\ \hline
B                                                                 & Service Recon                                                       & 20,982                   & 2,061,655                  & {[}8000, 88, 80, 8000, 8081, 80, 80{]}                                                                \\ \hline
C                                                                 & Basic Attacks \& Host Recon        & 113,407                  & 7,061,042                  & {[}445, 445, 445{]}                                                               \\ \hline
D                                                                 & Basic Attacks \& Host Recon        & 895                      & 68,065                     & {[}11390, 11390,  \dotso , 11390, 11390{]} \\ \hline
E                                                                 & Basic Attacks \& Host Recon & 285,651                  & 42,225,387                 & {[}23, 23, 2323{]}                                                                \\ \hline
F                                                                 & Complex Attacks & 43,305                   & 1,718,440                  & {[}9527, 9527, 9527, 5555, 5555, 5555{]}                                          \\ \hline
G                                                                 & Complex Attacks & 535                   & 7,270                    & {[}7550,7550,7547,7547,7547{]}                                          \\ \hline
H                                                                 & Complex Attacks & 43,305                   & 105,258                    & {[}7379, 7379, 5379, 5379, 6379, 6379{]}                                          \\ \hline
\end{tabular}}
\end{table*}

\subsubsection{Complex Attacks}
These clusters capture patterns involving multiple attack steps or vulnerabilities. This category of clusters often captures the most interesting attack patterns, and the most difficult to detect because of their low volume of traffic which hides them under the noise floor.
Moreover, without the proposed embedding approach, an attack which involves a sequence of ports would be clustered with the other Basic Attacks and thus the other ports in the attack would go unnoticed --making it impossible to distinguish from other attack campaigns.

We note that cluster $\Alph{ClusterExample}$ contains \%30 of all darknet sequences. This is because the cluster captures ports 23 and 2323 (which the embedding correctly associated together)
Both ports are used for Telnet, and they are connected with the Mirai botnet and are likely to appear in every darknet scanner from the last few years~\cite{heo2018}. 
\stepcounter{ClusterExample}

One example of this category occurred on 3/4/2019, where DANTE reported a new large cluster appeared ($\Alph{ClusterExample}$ in Figure~\ref{fig:overview}) consisting of two ports and originating from China and Brazil only. Here, 92\% of the src IPs sent the sequence $\{9527,9527,9527\}$, and 8\% of the src IPs sent $\{9527,9527,9527,5555,5555,5555\}$ and in the reverse order. This is a clear indication of one or more campaigns launched at the same time aiming to exploit a new vulnerability. Four days \textit{after} DANTE detected the attack, the attack was reported ~\cite{port9527} and related to a vulnerability in an IP-Camera (CVE-2017-11632).  Interestingly, none of the reports mentioned that the attackers were also targeting port 5555. This not only demonstrates how DANTE can provide threat intelligence, but how DANTE can detect ongoing attacks.
\stepcounter{ClusterExample}

\begin{figure}
\centering
\includegraphics[width=1\linewidth]{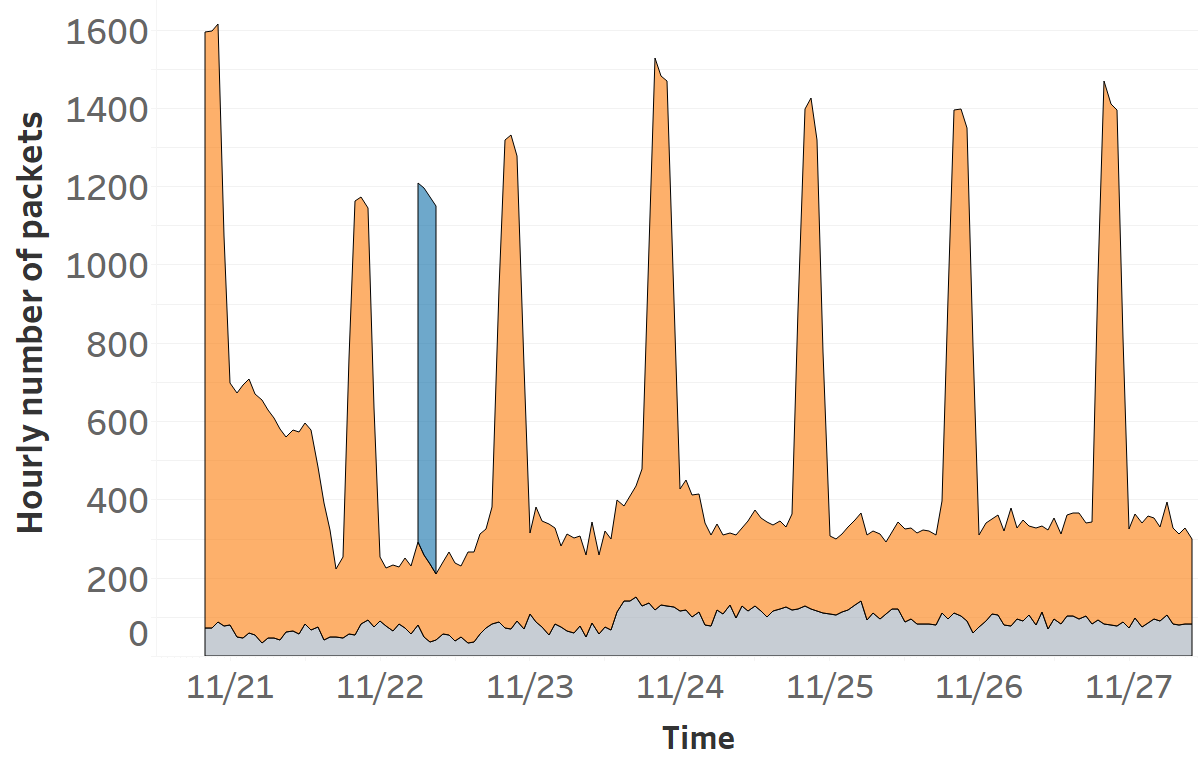}
\caption{
The number of packets that arrived to port 7547 during the week of 11/21/2018, where each color represents a different cluster.
The first cluster (orange) occurs each day at 10:00 AM GMT; the second cluster (gray) is small and occurs continuously during the week. 
DANTE detected the reported anomaly (blue) which occurred at 5:00 PM 11/22/18.  
}
\label{fig:port7547}
\end{figure}

A third example of this category can be seen in a cluster, $\Alph{ClusterExample}$, from 11/22/18, which was discovered by DANTE.
This cluster contains a pattern of scanning two specific ports, 7547 and 7550, which occurs on 11/22/2018 from 5 to 9 PM GMT. 
According to the Internet Storm Center (ISC) port search~\cite{van2008isc}, a free tool that monitors the level of malicious port activity, this is the most significant peak of activities for port 7550 in the past two years, however, to the best of our knowledge, there have been no reports of an attack that utilizes this port. 
The missing information in the reports could suggest a novel attack that utilize those two ports.
In addition, port 7547 appears to have a large number of packets arriving each day at 10:00 AM GMT, that were assigned to a different cluster. Unlike cluster $\Alph{ClusterExample}$, that cluster consists of port 7547 alone with no additional ports in the scan.
12/22/2018 is the only day when activity on this port peaks in a different hour (see Figure~\ref{fig:port7547}). 
There are reports~\cite{port7547} of a known Mirai botnet variant that uses port 7547 to exploit routers. 
Based on this report we attribute cluster \ $\Alph{ClusterExample}$ to Mirai activity.   
Since DANTE detected that the two ports are used interchangeably by the same src IPs, we suspect that there is a new vulnerability tested on port 7550 of routers.
At the time of writing of this article existence or absence of such vulnerability was not yet confirmed by any organization. \stepcounter{ClusterExample}

On 10/31/2018, DANTE generated an alert about a new attack (cluster $\Alph{ClusterExample}$) one minute after it began. This attack continued everyday from 11/15/18 until the 11/18/18. 
In the attack, 1,789 different network hosts sent exactly four packets to two or three of the ports: 5379, 6379, and 7379.
According to the ISC, this is the largest peak in the use of port 5379 and the third largest peak for port 7379, although these ports are considered unused.
We could not find any report online which identified these ports being used together in an attack. Identifying the source of these IPs can lead a CERT team to uncovering an ongoing campaign.

In addition, 44.8\% of the detected patterns (clusters) in this group reappeared on a later date (in some cases, one or two weeks later), sometimes with minor changes such as adding or removing some of the ports in the scan. 
Figure~\ref{fig:multiple_ports_gantt} presents a Gantt chart of those reoccurring clusters. 
In the case of the reoccurring patterns, DANTE did not send an alert regarding a new attack pattern but did report a new occurrence of the known pattern.

\stepcounter{ClusterExample}

\subsubsection{Noise and Outliers}
This category consisting of a single cluster, with the patterns that the DBScan algorithm defined as outliers.
Those patterns are port sequences that are not large enough, in terms of the number of src IPs, to become a new cluster.
The rationale behind this is that most of the traffic in this category is backscatter or misconfigurations packets, and thus does not represent a scan~\cite{Liu2014, harrop2005greynet}. 
Although some of the patterns can represent a scan, those kinds of patterns cannot indicate a trend due to their small volume. 
The size of this cluster is directly controlled by the minPts parameter in DBScan and can be changed at any time. 
As previously mentioned, we chose minPts to be 30 in order to create a reasonable number of clusters per day for a security expert to explore.

\begin{table*}[h!]
\centering
\caption{Top 20 biggest Batch 1 concepts according to DANTE and Ban et al. \label{freq_itemset}}
\begin{tabular}{|c|c|c|l|c|c|c|c|c|c|c|}
\cline{1-3} \cline{5-11}
\multicolumn{3}{|c|}{\textbf{Ban et al. \cite{ban2016}}} &  & \multicolumn{7}{c|}{\textbf{DANTE}}                                                        \\ \cline{1-3} \cline{5-11} 
Port 1      & Port 2      & Occur.         &  & Port 1               & Port 2 & Port 3 & Port 4 & Port 5 & Port 6             & Occur.     \\ \cline{1-3} \cline{5-11} 
23          &             & 5,783,748      &  & 23                   & 2323   &        &        &        &                    & 33,439,191 \\ \cline{1-3} \cline{5-11} 
2323        &             & 1,637,119      &  & 37215                &        &        &        &        &                    & 15,294,071 \\ \cline{1-3} \cline{5-11} 
2323        & 23          & 1,577,142      &  & 445                  &        &        &        &        &                    & 4,404,525  \\ \cline{1-3} \cline{5-11} 
445         &             & 982,528        &  & 80                 & 8001   & 8080   & 8081   & 8088   & and 10 other ports & 3,699,342  \\ \cline{1-3} \cline{5-11} 
80          &             & 886,540        &  & 80                   & 8080   & 433    &        &        &                    & 3,355,387  \\ \cline{1-3} \cline{5-11} 
37215       &             & 672,897        &  & 22                   &        &        &        &        &                    & 2,697,434  \\ \cline{1-3} \cline{5-11} 
8080        &             & 539,976        &  & 8080                 & 8081   &        &        &        &                    & 2,403,517  \\ \cline{1-3} \cline{5-11} 
8081        &             & 383,209        &  & 3389                 & 3398   & 9833   & 3839   & 8933   &                    & 1,383,441  \\ \cline{1-3} \cline{5-11} 
5555        &             & 322,826        &  & 5555                 &        &        &        &        &                    & 937,267    \\ \cline{1-3} \cline{5-11} 
8080        & 80          & 277,633        &  & 1433                 & 3413   &        &        &        &                    & 850,124    \\ \cline{1-3} \cline{5-11} 
37215       & 23          & 244,212        &  & 1235 different ports &        &        &        &        &                    & 798,760    \\ \cline{1-3} \cline{5-11} 
81          &             & 187,815        &  & 80                   & 8080   &        &        &        &                    & 615,490    \\ \cline{1-3} \cline{5-11} 
3389        &             & 178,766        &  & 3389                 & 139    & 445    & 11211  &        &                    & 519,732    \\ \cline{1-3} \cline{5-11} 
8443        &             & 158,892        &  & 21                   &        &        &        &        &                    & 482,898    \\ \cline{1-3} \cline{5-11} 
32764       &             & 132,361        &  & 81                   &        &        &        &        &                    & 463,355    \\ \cline{1-3} \cline{5-11} 
1433        &             & 130,363        &  & 1252 different ports &        &        &        &        &                    & 344,761    \\ \cline{1-3} \cline{5-11} 
9000        &             & 129,044        &  & 443                  &        &        &        &        &                    & 343,922    \\ \cline{1-3} \cline{5-11} 
22          &             & 122,518        &  & 8080                 & 8000   & 8333   &        &        & and 44 other ports & 237,758    \\ \cline{1-3} \cline{5-11} 
80          & 23          & 106,023        &  & 123                  & 47808  & 161    & 1883   & 8883   & 501                & 195,431    \\ \cline{1-3} \cline{5-11} 
21          &             & 104181         &  & 5900                 & 5901   & 5431   &        &        &                    & 150,234    \\ \cline{1-3} \cline{5-11} 
\end{tabular}

\end{table*}

\subsection{Comparative Discussion}
In the previous sections, we demonstrated that DANTE can detect new and reoccurring attack patterns. 
As mentioned in section \ref{sec:relatedWork}, methods such as \cite{thonnard2008, ban2015, ban2016, Liu2014} can detect port scanning (access to many different ports per host), and methods such as \cite{Liu2014, thonnard2008, inoue2008incident, kao2015predictive} can observe traffic spikes on individual port numbers. However, DANTE is able to detect both port scans and port spikes while distinguishing between them. 
Furthermore, because DANTE uses Word2vec, it is able to cluster meaningful patterns consisting of different ports that have the same meaning/intent. For example, cluster B consists of port sequences which revolve around a specific type of service. Therefore, DANTE can detect both simple and complex patterns.

The greatest advantage of DANTE is its ability to detect patterns from the Complex Attacks category (e.g., CVE-2017-11632). This family, consisting of multiple ports per sequence, captures many well known and dangerous malwares. Table \ref{malware_ports} exemplifies this claim by listing some of these malwares along with the ports which they access during their lifetime. To the best of our knowledge, there are no other works which can detect these kinds of patterns as quickly and accurately as DANTE.  Concretely, \cite{thonnard2008, Liu2014, coudriau2016topological, inoue2008incident, kao2015predictive} observe traffic on individual ports. Although \cite{ban2015, ban2016} finds the connection between ports, it ignores the meaning and the content of the sequences. Another example is that \cite{thonnard2008} would not be able to detect cluster G that consist of ports 7547 and 7550 as it looks for common temporal activities spike of ports. While port 9537 had a significant spike at this date, there were no spike in the usage of port 5555 as so the two ports would not be clustered together. 
In contrast, DANTE uses the embedding of the ports to find the connection between them in near real time, and found that this is the first occurrence of those two ports together and issued an alert to the CERT within minutes, four full days before the first online report.

\subsection{Comparison with Ban \textit{et al.}~\cite{ban2016}}

To evaluate the contribution of using Word2vec as a means for mining meaningful patterns, we compare our embedding approach to Ban \textit{et al.}~\cite{ban2016}. 
Ban \textit{et al.} used a frequent pattern mining (FPM) algorithm called FP-growth on darknet traffic to group similar ports together. 
By using FPM, Ban \textit{et al.} was able to find subsets of ports that occur together in the data.
To the best of our knowledge, this is the most recent work that mines patterns in darknet traffic to find new concepts.
We applied their method on Batch 1 using the same time windows.
By using the same parameters and aggregations, we were able to extract the groups of ports and their number of occurrences (support). 
The top 20 largest subsets of both FPM and DANTE are presented in Table~\ref{freq_itemset}.


While FPM detects different subsets of ports, it can not identify when different subsets are jointly used in novel attack patterns. 
This can be seen in cluster F, where DANTE clusters two different sets together, the set [9527] and the set [5555, 9527]. 
This clustering led to the discovery of port 5555 being used against Wireless IP Camera 360 devices in vulnerability CVE-2017-11632. 
In addition, Ban \textit{et al.}'s method omits all sequences with more than six different ports, while these patterns are important in detecting malicious service reconnaissance.
For example, in cluster A DANTE discovered a malicious actor who was copying Censys's port scanning behaviors.

By using the semantic embedding of ports, DANTE learns that different ports which behave similarly (i.e., using the same service) should be considered close in the embedding space. 
That is why in cluster E, DANTE clustered together appearances of port 23 and port 2323 (both used by Telnet). 
On the other hand, the FPM algorithm could only create a set for every permutation of the ports although they all capture the same concept (see the three biggest FPM sets in Table~\ref{freq_itemset}).

Moreover, some ports typically appear in attack sequences and rarely appear alone. This means that there is semantic information which can be learned about these ports, such as the intent and how it is reflected on the other ports in the sequence. 
For example, in Table \ref{freq_itemset}, port 80 (HTTP) is a single concept according to FPM due to the FP-growth process. In contrast, DANTE re-associates port 80 with other patterns. 
This can be seen in the top 20 largest concepts where port 80 appears in six different concepts but never appears by itself; in FPM, the singleton [80] is the fifth largest subset.

Lastly, FP-growth creates a large number of sets, even when setting the minimum support parameter to be a relatively large value (1,819 sets with minimum support of 1,000 for a period of six weeks). 
The resulting number of sets is impractical for a CERT analyst to investigate.
An advantage of using DANTE is that only a reasonable number of concepts with high importance are identified per day (average of four a day, as shown in Figure~\ref{fig:new_clusters_count}). At the same time, DANTE discovers small yet significant patterns that might represent dangerous attacks (such as cluster G).

\section{Conclusion and Future Work}
\label{sec:conclusionAndFutureWork}

By mining darknet traffic, analysts can get get frequent reports on on-going and new merging threats facing their network. In this paper, we presented DANTE: a framework which enables network service providers to mine threat intelligence from massive darknet traffic streams. The framework provides (1) a novel method for representing a set of targeted ports as an embedding which captures the intent of the attack, and (2) a system for clustering, tracking, and detecting old and new attack patterns. 

By evaluating DANTE on real darknet traffic, we were able to confirm that DANTE can track and detect reoccurring and new attack patterns.
For example, the system detected a reported attack on IP cameras on the 3/4/2019. Furthermore, we discovered some attacks which have not been reported. For example, the patterns involving ports [5379, 6379, 7379] which appeared on the 10/31/2018. 
We also compared DANTE to a well-known darknet mining algorithm and found that DANTE produced higher quality and results with practical outputs. 

DANTE is currently being deployed in Deutsche Telekom's networks to provide their CERT with better threat intelligence.

As future work, we plan to improve DANTE by learning a semantic relationship between the port-embeddings and other features such as geo-location, ttl, packet size, and others.

\begin{table}[]
\caption{Different malwares and the ports they are using for infection and communication. \label{malware_ports}}

\resizebox{\linewidth}{!}{%

\begin{tabular}{|c|c|c|}
\hline
\textbf{Botnet} & \textbf{Ports}                                                                    & \textbf{Phase} \\ \hline
Cryptojacking   & \begin{tabular}[c]{@{}c@{}}6379, 2375, 2376\end{tabular}                      & Infection      \\ \hline
Wicked          & \begin{tabular}[c]{@{}c@{}}8080, 8443, 80, 81\end{tabular}                    & Infection      \\ \hline
MIRAI           & 23, 2323                                                                          & Infection      \\ \hline
CONFICKER       & 445                                                                               & Infection      \\ \hline
EMOTET          & \begin{tabular}[c]{@{}c@{}}80, 990, 8090,\\   50000, 8080, 7080, 443\end{tabular} & Post-infection \\ \hline
CERBER          & 80, 6892                                                                          & Post-infection \\ \hline
KOVTER          & 43, 8080                                                                          & Post-infection \\ \hline
QAKBOT          & 443, 65400 , 2222 , 21, 41947                                                     & Post-infection \\ \hline
Cryptojacking   & \begin{tabular}[c]{@{}c@{}}22, 7878, 2375, 2376\end{tabular}                  & Post-infection \\ \hline
GH0ST           & \begin{tabular}[c]{@{}c@{}}2011, 2013, 800\end{tabular}                       & Post-infection \\ \hline
NANOCORE        & 3365                                                                              & Post-infection \\ \hline
\end{tabular}}

\end{table}

We hope that the framework described in this paper, the embedding and clustering techniques, will assist researchers and the industry in better securing the Internet.

\section*{Acknowledgements} 
\label{sec:Acknowledgements}
We would like to thank Nadav Maman for his help in implementing this work in the Spark environment. 

\small{
\bibliographystyle{IEEEtran}
\bibliography{biblio}

\begin{thebibliography}{10}
\providecommand{\url}[1]{#1}
\csname url@samestyle\endcsname
\providecommand{\newblock}{\relax}
\providecommand{\bibinfo}[2]{#2}
\providecommand{\BIBentrySTDinterwordspacing}{\spaceskip=0pt\relax}
\providecommand{\BIBentryALTinterwordstretchfactor}{4}
\providecommand{\BIBentryALTinterwordspacing}{\spaceskip=\fontdimen2\font plus
\BIBentryALTinterwordstretchfactor\fontdimen3\font minus
  \fontdimen4\font\relax}
\providecommand{\BIBforeignlanguage}[2]{{%
\expandafter\ifx\csname l@#1\endcsname\relax
\typeout{** WARNING: IEEEtran.bst: No hyphenation pattern has been}%
\typeout{** loaded for the language `#1'. Using the pattern for}%
\typeout{** the default language instead.}%
\else
\language=\csname l@#1\endcsname
\fi
#2}}
\providecommand{\BIBdecl}{\relax}
\BIBdecl

\bibitem{bailey2005internet}
M.~Bailey, E.~Cooke, F.~Jahanian, J.~Nazario, D.~Watson \emph{et~al.}, ``The
  internet motion sensor-a distributed blackhole monitoring system.'' in
  \emph{NDSS}, 2005.

\bibitem{bailey2006practical}
M.~Bailey, E.~Cooke, F.~Jahanian, A.~Myrick, and S.~Sinha, ``Practical darknet
  measurement,'' in \emph{2006 40th Annual Conference on Information Sciences
  and Systems}.\hskip 1em plus 0.5em minus 0.4em\relax IEEE, 2006, pp.
  1496--1501.

\bibitem{mairh2011honeypot}
A.~Mairh, D.~Barik, K.~Verma, and D.~Jena, ``Honeypot in network security: a
  survey,'' in \emph{Proceedings of the 2011 international conference on
  communication, computing \& security}.\hskip 1em plus 0.5em minus 0.4em\relax
  ACM, 2011, pp. 600--605.

\bibitem{bringer2012honeypot}
M.~L. Bringer, C.~A. Chelmecki, and H.~Fujinoki, ``A survey: Recent advances
  and future trends in honeypot research,'' \emph{International Journal of
  Computer Network and Information Security}, vol.~4, no.~10, p.~63, 2012.

\bibitem{Pang2016}
\BIBentryALTinterwordspacing
S.~Pang, D.~Komosny, {Lei Zhu}, R.~Zhang, A.~Sarrafzadeh, {Tao Ban}, and
  {Daisuke Inoue}, ``{Malicious Events Grouping via Behavior Based Darknet
  Traffic Flow Analysis},'' \emph{Wireless Personal Communications}, vol.~96.
  [Online]. Available:
  \url{https://link.springer.com/content/pdf/10.1007{\%}2Fs11277-016-3744-4.pdf}
\BIBentrySTDinterwordspacing

\bibitem{Wustrow2010}
\BIBentryALTinterwordspacing
E.~Wustrow, M.~Karir, M.~Bailey, F.~Jahanian, and G.~Huston, ``Internet
  background radiation revisited,'' in \emph{Proceedings of the 10th ACM
  SIGCOMM Conference on Internet Measurement}, ser. IMC '10.\hskip 1em plus
  0.5em minus 0.4em\relax New York, NY, USA: ACM, 2010, pp. 62--74. [Online].
  Available: \url{http://doi.acm.org/10.1145/1879141.1879149}
\BIBentrySTDinterwordspacing

\bibitem{ban2012}
T.~Ban, L.~Zhu, J.~Shimamura, S.~Pang, D.~Inoue, and K.~Nakao, ``Behavior
  analysis of long-term cyber attacks in the darknet,'' in \emph{International
  Conference on Neural Information Processing}.\hskip 1em plus 0.5em minus
  0.4em\relax Springer, 2012, pp. 620--628.

\bibitem{Pa2016}
Y.~M.~P. Pa, S.~Suzuki, K.~Yoshioka, T.~Matsumoto, T.~Kasama, and C.~Rossow,
  ``Iotpot: A novel honeypot for revealing current iot threats,'' \emph{Journal
  of Information Processing}, vol.~24, no.~3, pp. 522--533, 2016.

\bibitem{Fachkha2015}
\BIBentryALTinterwordspacing
C.~Fachkha, E.~Bou-Harb, and M.~Debbabi, ``{Inferring distributed reflection
  denial of service attacks from darknet},'' \emph{Computer Communications},
  vol.~62, pp. 59--71, may 2015. [Online]. Available:
  \url{https://www.sciencedirect.com/science/article/pii/S0140366415000316}
\BIBentrySTDinterwordspacing

\bibitem{Bou-Harb2015}
\BIBentryALTinterwordspacing
E.~Bou-Harb, M.~Debbabi, and C.~Assi, ``{A Time Series Approach for Inferring
  Orchestrated Probing Campaigns by Analyzing Darknet Traffic},'' in \emph{2015
  10th International Conference on Availability, Reliability and
  Security}.\hskip 1em plus 0.5em minus 0.4em\relax IEEE, aug 2015, pp.
  180--185. [Online]. Available:
  \url{http://ieeexplore.ieee.org/document/7299912/}
\BIBentrySTDinterwordspacing

\bibitem{singhal2017security}
A.~Singhal and X.~Ou, ``Security risk analysis of enterprise networks using
  probabilistic attack graphs,'' in \emph{Network Security Metrics}.\hskip 1em
  plus 0.5em minus 0.4em\relax Springer, 2017, pp. 53--73.

\bibitem{mikolov2013}
\BIBentryALTinterwordspacing
T.~Mikolov, K.~Chen, G.~Corrado, and J.~Dean, ``Efficient estimation of word
  representations in vector space,'' \emph{CoRR}, vol. abs/1301.3781, 2013.
  [Online]. Available: \url{http://arxiv.org/abs/1301.3781}
\BIBentrySTDinterwordspacing

\bibitem{ban2016}
T.~Ban, S.~Pang, M.~Eto, D.~Inoue, K.~Nakao, and R.~Huang, ``Towards early
  detection of novel attack patterns through the lens of a large-scale
  darknet,'' in \emph{2016 Intl IEEE Conferences on Ubiquitous Intelligence
  Computing, Advanced and Trusted Computing, Scalable Computing and
  Communications, Cloud and Big Data Computing, Internet of People, and Smart
  World Congress (UIC/ATC/ScalCom/CBDCom/IoP/SmartWorld)}, July 2016, pp.
  341--349.

\bibitem{ban2015}
T.~Ban, M.~Eto, S.~Guo, D.~Inoue, K.~Nakao, and R.~Huang, ``A study on
  association rule mining of darknet big data,'' in \emph{2015 International
  Joint Conference on Neural Networks (IJCNN)}, July 2015, pp. 1--7.

\bibitem{Ban2017}
T.~Ban, L.~Zhu, J.~Shimamura, S.~Pang, D.~Inoue, and K.~Nakao, ``Detection of
  botnet activities through the lens of a large-scale darknet,'' in
  \emph{Neural Information Processing}.\hskip 1em plus 0.5em minus 0.4em\relax
  Cham: Springer International Publishing, 2017, pp. 442--451.

\bibitem{Furutani2014}
\BIBentryALTinterwordspacing
N.~Furutani, T.~Ban, J.~Nakazato, J.~Shimamura, J.~Kitazono, and S.~Ozawa,
  ``{Detection of DDoS Backscatter Based on Traffic Features of Darknet TCP
  Packets},'' in \emph{2014 Ninth Asia Joint Conference on Information
  Security}.\hskip 1em plus 0.5em minus 0.4em\relax IEEE, sep 2014, pp. 39--43.
  [Online]. Available: \url{http://ieeexplore.ieee.org/document/7023237/}
\BIBentrySTDinterwordspacing

\bibitem{Liu2014}
\BIBentryALTinterwordspacing
J.~Liu and K.~Fukuda, ``{Towards a taxonomy of darknet traffic},'' in
  \emph{2014 International Wireless Communications and Mobile Computing
  Conference (IWCMC)}.\hskip 1em plus 0.5em minus 0.4em\relax IEEE, aug 2014,
  pp. 37--43. [Online]. Available:
  \url{http://ieeexplore.ieee.org/lpdocs/epic03/wrapper.htm?arnumber=6906329}
\BIBentrySTDinterwordspacing

\bibitem{Choi2013}
\BIBentryALTinterwordspacing
S.-s. Choi, J.~Song, S.~Kim, and S.~Kim, ``{A model of analyzing cyber threats
  trend and tracing potential attackers based on darknet traffic},''
  \emph{Security and Communication Networks}, vol.~7, no.~10, pp. n/a--n/a, may
  2013. [Online]. Available: \url{http://doi.wiley.com/10.1002/sec.796}
\BIBentrySTDinterwordspacing

\bibitem{skrjanc2017}
I.~Škrjanc, S.~Ozawa, D.~Dovžan, B.~Tao, J.~Nakazato, and J.~Shimamura,
  ``Evolving cauchy possibilistic clustering and its application to large-scale
  cyberattack monitoring,'' in \emph{2017 IEEE Symposium Series on
  Computational Intelligence (SSCI)}, Nov 2017, pp. 1--7.

\bibitem{thonnard2008}
O.~Thonnard and M.~Dacier, ``A framework for attack patterns' discovery in
  honeynet data,'' \emph{digital investigation}, vol.~5, pp. S128--S139, 2008.

\bibitem{coudriau2016topological}
M.~{Coudriau}, A.~{Lahmadi}, and J.~{François}, ``Topological analysis and
  visualisation of network monitoring data: Darknet case study,'' in \emph{2016
  IEEE International Workshop on Information Forensics and Security (WIFS)},
  Dec 2016, pp. 1--6.

\bibitem{Lagraa2017}
\BIBentryALTinterwordspacing
S.~Lagraa, J.~Francois, A.~Lahmadi, M.~Miner, C.~Hammerschmidt, and R.~State,
  ``{BotGM: Unsupervised graph mining to detect botnets in traffic flows},'' in
  \emph{2017 1st Cyber Security in Networking Conference (CSNet)}.\hskip 1em
  plus 0.5em minus 0.4em\relax IEEE, oct 2017, pp. 1--8. [Online]. Available:
  \url{http://ieeexplore.ieee.org/document/8241990/}
\BIBentrySTDinterwordspacing

\bibitem{Owezarski2015}
\BIBentryALTinterwordspacing
P.~Owezarski, ``{A Near Real-Time Algorithm for Autonomous Identification and
  Characterization of Honeypot Attacks},'' Tech. Rep., 2015. [Online].
  Available: \url{https://hal.archives-ouvertes.fr/hal-01112926}
\BIBentrySTDinterwordspacing

\bibitem{Casas2012}
\BIBentryALTinterwordspacing
P.~Casas, J.~Mazel, and P.~Owezarski, ``{Unsupervised Network Intrusion
  Detection Systems: Detecting the Unknown without Knowledge},'' \emph{Computer
  Communications}, vol.~35, pp. 772--783, 2012. [Online]. Available:
  \url{https://ac.els-cdn.com/S0140366412000266/1-s2.0-S0140366412000266-main.pdf?{\_}tid=78066dd0-7890-4283-91d5-24d68196af9a{\&}acdnat=1537171232{\_}8b26ff3454a5d96541ba94b7689009a0}
\BIBentrySTDinterwordspacing

\bibitem{Corchado2010}
\BIBentryALTinterwordspacing
E.~Corchado and {\'{A}}.~Herrero, ``{Neural visualization of network traffic
  data for intrusion detection},'' \emph{Applied Soft Computing Journal},
  vol.~11, pp. 2042--2056, 2010. [Online]. Available:
  \url{https://ac.els-cdn.com/S1568494610001559/1-s2.0-S1568494610001559-main.pdf?{\_}tid=05fc7848-1a31-45c0-a0eb-36d924e6aaba{\&}acdnat=1537171402{\_}7532d430741abdc09424106711827705}
\BIBentrySTDinterwordspacing

\bibitem{Bartos2016}
\BIBentryALTinterwordspacing
K.~Bartos, M.~Sofka, and V.~Franc, \emph{{Optimized Invariant Representation of
  Network Traffic for Detecting Unseen Malware Variants}}, 2016. [Online].
  Available:
  \url{https://www.usenix.org/conference/usenixsecurity16/technical-sessions/presentation/bartos}
\BIBentrySTDinterwordspacing

\bibitem{Zhang2016}
\BIBentryALTinterwordspacing
J.~Zhang, Y.~Tong, and T.~Qin, ``{Traffic features extraction and clustering
  analysis for abnormal behavior detection},'' in \emph{Proceedings of the 2016
  International Conference on Intelligent Information Processing - ICIIP
  '16}.\hskip 1em plus 0.5em minus 0.4em\relax New York, New York, USA: ACM
  Press, 2016, pp. 1--6. [Online]. Available:
  \url{http://dl.acm.org/citation.cfm?doid=3028842.3028867}
\BIBentrySTDinterwordspacing

\bibitem{ester1996density}
M.~Ester, H.-P. Kriegel, J.~Sander, X.~Xu \emph{et~al.}, ``A density-based
  algorithm for discovering clusters in large spatial databases with noise.''
  in \emph{Kdd}, vol.~96, no.~34, 1996, pp. 226--231.

\bibitem{guha2000}
S.~Guha, N.~Mishra, R.~Motwani, and L.~O'Callaghan, ``Clustering data
  streams,'' in \emph{Foundations of computer science, 2000. proceedings. 41st
  annual symposium on}.\hskip 1em plus 0.5em minus 0.4em\relax IEEE, 2000, pp.
  359--366.

\bibitem{ester1998}
M.~Ester, H.-P. Kriegel, J.~Sander, M.~Wimmer, and X.~Xu, ``Incremental
  clustering for mining in a data warehousing environment,'' in \emph{VLDB},
  vol.~98.\hskip 1em plus 0.5em minus 0.4em\relax Citeseer, 1998, pp. 323--333.

\bibitem{cao2006}
F.~Cao, M.~Estert, W.~Qian, and A.~Zhou, ``Density-based clustering over an
  evolving data stream with noise,'' in \emph{Proceedings of the 2006 SIAM
  international conference on data mining}.\hskip 1em plus 0.5em minus
  0.4em\relax SIAM, 2006, pp. 328--339.

\bibitem{Carnein2019}
\BIBentryALTinterwordspacing
M.~Carnein and H.~Trautmann, ``Optimizing data stream representation: An
  extensive survey on stream clustering algorithms,'' \emph{Business {\&}
  Information Systems Engineering}, Jan 2019. [Online]. Available:
  \url{https://doi.org/10.1007/s12599-019-00576-5}
\BIBentrySTDinterwordspacing

\bibitem{faria2016novelty}
E.~R. Faria, I.~J. Gon{\c{c}}alves, A.~C. de~Carvalho, and J.~Gama, ``Novelty
  detection in data streams,'' \emph{Artificial Intelligence Review}, vol.~45,
  no.~2, pp. 235--269, 2016.

\bibitem{misp2016}
C.~Wagner, A.~Dulaunoy, G.~Wagener, and A.~Iklody, ``Misp: The design and
  implementation of a collaborative threat intelligence sharing platform,'' in
  \emph{Proceedings of the 2016 ACM on Workshop on Information Sharing and
  Collaborative Security}.\hskip 1em plus 0.5em minus 0.4em\relax ACM, 2016,
  pp. 49--56.

\bibitem{wieting2015towards}
J.~Wieting, M.~Bansal, K.~Gimpel, and K.~Livescu, ``Towards universal
  paraphrastic sentence embeddings,'' \emph{arXiv preprint arXiv:1511.08198},
  2015.

\bibitem{maaten2008visualizing}
L.~v.~d. Maaten and G.~Hinton, ``Visualizing data using t-sne,'' \emph{Journal
  of machine learning research}, vol.~9, no. Nov, pp. 2579--2605, 2008.

\bibitem{harrop2005greynet}
W.~Harrop and G.~Armitage, ``Defining and evaluating greynets (sparse
  darknets),'' in \emph{The IEEE Conference on Local Computer Networks 30th
  Anniversary (LCN'05) l}.\hskip 1em plus 0.5em minus 0.4em\relax IEEE, 2005,
  pp. 344--350.

\bibitem{Durumeric2014}
\BIBentryALTinterwordspacing
Z.~Durumeric, M.~Bailey, and J.~A. Halderman, ``An internet-wide view of
  internet-wide scanning,'' in \emph{Proceedings of the 23rd USENIX Conference
  on Security Symposium}, ser. SEC'14.\hskip 1em plus 0.5em minus 0.4em\relax
  Berkeley, CA, USA: USENIX Association, 2014, pp. 65--78. [Online]. Available:
  \url{http://dl.acm.org/citation.cfm?id=2671225.2671230}
\BIBentrySTDinterwordspacing

\bibitem{heo2018}
H.~Heo and S.~Shin, ``Who is knocking on the telnet port: A large-scale
  empirical study of network scanning,'' in \emph{Proceedings of the 2018 on
  Asia Conference on Computer and Communications Security}.\hskip 1em plus
  0.5em minus 0.4em\relax ACM, 2018, pp. 625--636.

\bibitem{port9527}
\BIBentryALTinterwordspacing
S.~Nichols. (2019) Fbi warns of sim-swap scams, ibm finds holes in visitor
  software, 13-year-old girl charged over javascript prank. [Online].
  Available:
  \url{\url{https://www.theregister.co.uk/2019/03/09/security\_roundup\_080319}}
\BIBentrySTDinterwordspacing

\bibitem{van2008isc}
M.~Van~Horenbeeck, ``The sans internet storm center,'' in \emph{2008 WOMBAT
  Workshop on Information Security Threats Data Collection and Sharing}.\hskip
  1em plus 0.5em minus 0.4em\relax IEEE, 2008, pp. 17--23.

\bibitem{port7547}
\BIBentryALTinterwordspacing
J.~Ullrich. (2016) Port 7547 soap remote code execution attack against dsl
  modems. [Online]. Available:
  \url{https://isc.sans.edu/diary/Port+7547+SOAP+Remote+Code+
  Execution+Attack+Against+DSL+Modems/21759}
\BIBentrySTDinterwordspacing

\bibitem{inoue2008incident}
D.~Inoue, K.~Yoshioka, M.~Eto, M.~Yamagata, E.~Nishino, J.~Takeuchi,
  K.~Ohkouchi, and K.~Nakao, ``An incident analysis system nicter and its
  analysis engines based on data mining techniques,'' in \emph{International
  Conference on Neural Information Processing}.\hskip 1em plus 0.5em minus
  0.4em\relax Springer, 2008, pp. 579--586.

\bibitem{kao2015predictive}
C.-N. Kao, Y.-C. Chang, N.-F. Huang, I.-J. Liao, R.-T. Liu, H.-W. Hung
  \emph{et~al.}, ``A predictive zero-day network defense using long-term
  port-scan recording,'' in \emph{2015 IEEE Conference on Communications and
  Network Security (CNS)}.\hskip 1em plus 0.5em minus 0.4em\relax IEEE, 2015,
  pp. 695--696.

\end{thebibliography}

}
\normalsize

{\vskip 12pt}
\noindent
\end{document}